\documentclass{aa}  

\usepackage{graphicx}
\usepackage{empheq}
\usepackage{txfonts}
\usepackage{soul}
\begin{document} 
   \title{Creep Tide Model for the 3-Body Problem}

   \subtitle{The rotational evolution of a circumbinary planet }

   \author{F. A. Zoppetti \inst{1,2},
          H. Folonier \inst{3,4},
          A. M. Leiva \inst{1} \and 
          C. Beaugé \inst{1,2} 
          }

   \institute{Observatorio Astronómico de Córdoba, Universidad Nacional de Córdoba, Laprida 854, Córdoba X5000BGR, Argentina
\\ \email{federico.zopetti@unc.edu.ar}
         \and
             CONICET, Instituto de Astronomía Teórica y Experimental, Laprida 854, Córdoba X5000BGR, Argentina 
         \and   Instituto de Astronomia Geofísica e Ciências Atmosféricas, Universidade de São Paulo, SP 05508-090, Brazil  
}

   \date{}

 

  \abstract
  {We present a tidal model for treating the rotational evolution in the general three-body problem with arbitrary viscosities, in which all the masses are considered to be extended and all the tidal interactions between pairs are taken into account. Based on the creep tide theory, we present the set of differential equations that describes the rotational evolution of each body, in a formalism that is easily extensible to the {\it N} tidally-interacting body problem. We apply our model to the case of a circumbinary planet and use a Kepler-38 like binary system as a working example. We find that, in this low planetary eccentricity case, the most likely final stationary rotation state is the 1:1 spin-orbit resonance, considering an arbitrary planetary viscosity inside the estimated range for the solar system planets. The timescales for reaching the equilibrium state are expected to be $\sim$ Myrs for stiff bodies but can be longer than the age of the system for planets with a large gaseous component. We derive analytical expressions for the mean rotational stationary state, based on high-order power series of the semimajor axes ratio $a_1/a_2$ and low-order expansions of the eccentricities. These are found to reproduce very accurately the mean behaviour of the low-eccentric numerical integrations for arbitrary planetary relaxation factors, and up to $a_1/a_2 \sim 0.4$. Our analytical model is used to predict the stationary rotation of the {\it{Kepler}} circumbinary planets and find that most of them are probably rotating in a sub-synchronous state, although the synchrony shift is much less important than the one estimated in \cite{Zoppetti2019,Zoppetti2020}. We present a comparison of our results with those obtained with the Constant Time Lag and find that, unlike what we assumed in our previous works, the cross torques have a non-negligible net secular contribution, and must be taken into account when computing the tides over each body in an {\it N}-extended-body system from an arbitrary reference frame. These torques are naturally taken into account in the creep theory. In addition to this, the latter formalism considers more realistic rheology that proved to reduce to the Constant Time Lag model in the gaseous limit and also allows to study several additional relevant physical phenomena.} 
  


   \maketitle
%

\section{Introduction}

It is well known that the rotational evolution of extended bodies in compact systems is strongly affected by tides.
In the two-body problem (2BP), tides tend to drive the spins to a stationary state characterized by a spin-orbit resonance. In particular, tidal models following the Darwin theory predict the existence of a stationary solution synchronous with the mean motion, when the two bodies move in circular orbits, or super-synchronous otherwise \citep[e.g.][]{FerrazMello2008}. However, this prediction has not been confirmed by the observations of satellites rotation rates, as in the case of Titan (e.g. Meriggiola 2012), Europa \citep[e.g.][]{Hurford2007} and the moon itself.

Several attempts were made to explain these differences, either considering an extra torque caused by the existence of a permanent deformation, or by modifying the tidal theories. In particular, \cite{FerrazMello2013,FerrazMello2015} proposed a theory of tidal interactions based on an approximate solution of the Navier–Stokes equation, where the capture in spin–orbit resonances can be explained without the assumption of an extra torque. This result was also reported by \cite{Makarov2013} and confirmed by \cite{Correia2014} with a model based on a Maxwell viscoelastic rheology. 

Regarding the general three-body problem (3BP), a model that takes into account all tidal interactions between pairs and valid for bodies with arbitrary viscosities is still missing. Instead, this problem is usually addressed by neglecting some of the tidal interactions. For example, in the circumbinary problem in which the planetary mass is expected to be much smaller than the stellar masses, the planetary tides induced on the stellar components are commonly omitted \cite[e.g.][]{Correia2016}. Although the appropriateness of these approximations is not under debate, building a general model valid for any mass ratio between pairs, allows extending the results to problems with different hierarchies, such as triple star systems and exomoons host planets close to the central star.

In \cite{Zoppetti2019} we presented a self-consistent tidal model for the planar and non-oblique 3BP, using a weak friction Constant Time Lag (CTL) model, based on the expressions of the tidal forces and torques of the classical work of \cite{Mignard1979}. We applied the model to the circumbinary (CB) case and found that for systems like those discovered by the {\it{Kepler mission}}, the planet acquired stationary spin rates on timescales of $\sim$ Myrs (depending on the dissipation factor $Q_2$) and, curiously, in sub-synchronous solutions \citep{Zoppetti2020}. The sub-synchronous of the planetary spin was found to be the result of the tidal interaction with the secondary central star and proportional to its mass, while the planetary eccentricity tended to generate super-synchronous states, as in the 2BP. However, the goal of building a general tidal model in the 3BP has not yet been fulfilled since the weak-friction regime of the CTL models has shown to be only valid for bodies with a large gaseous component \citep{Efroimsky2012,Efroimsky2015}.

In this work, we present and discuss a self-consistent model for treating the tides in the 3BP, in the framework of the Newtonian creep tidal model \cite{FerrazMello2013}. We focus our attention on the rotational evolution of the bodies and apply the results to the case of a CB planet. In particular, to allow for comparison with previous results, we employ a Kepler-38 like system \citep{Orosz2012}. However, a wider range of system parameters as well as different initial orbital elements are also explored.

This paper is organized as follows. In Section 2 we present the tidal model by computing the equilibrium figures of the bodies and deriving differential equations from the creep equation to model the real shapes and orientations for each one. Once the real figures are obtained, the rotational evolution is calculated from the reaction torques on the two companions. Section 3 presents a series of numerical integrations of the shape, orientation, and spin evolution equations for a CB planet in a Kepler-38 system, considering different planetary internal viscosities. In Section 4, we construct analytical expressions for the mean values of the rotational stationary state of the CB planet from a Jacobian reference frame and compare these predictions with numerical solutions. In addition to this, we estimate the equilibrium rotation of the {\it{Kepler}} CB planets. In Section 5, we compare our results with those obtained in \cite{Zoppetti2019,Zoppetti2020}, and present an improved formalism that takes into account the cross torques in the CTL model. Finally, Section 6 summarises our main results and discusses their implications.

\section{The Creep Tidal Model for the 3BP}

\begin{figure}[h!]
\centering
\includegraphics*[width=0.8\columnwidth]{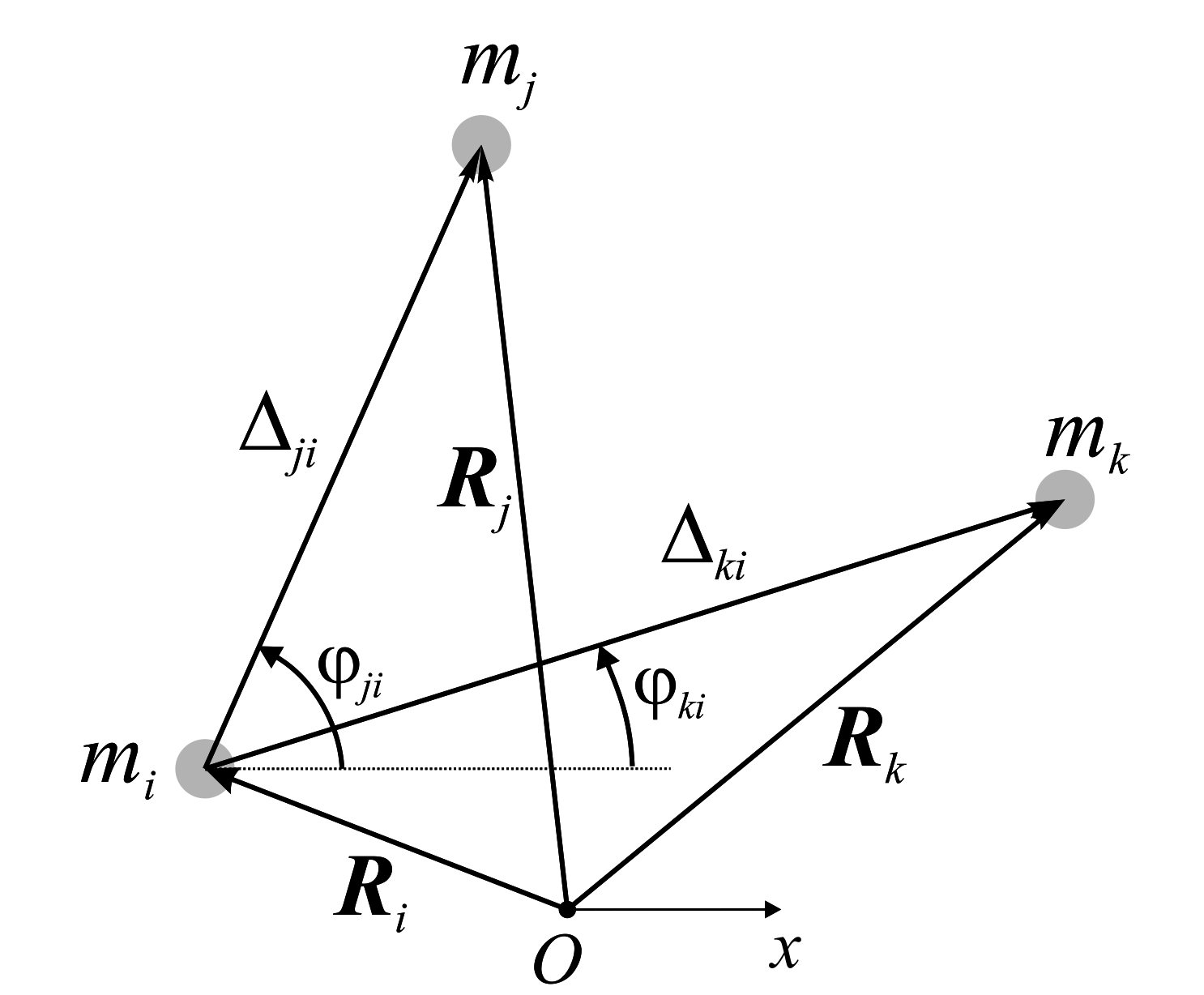}
\caption{The three-body problem from an inertial frame and from a $m_i$-relative frame.}
\label{3c}
\end{figure}

Let us consider the extended three-body problem with masses $m_i$, $m_j$ and $m_k$ and respective mean radius $\mathcal{R}_i$, $\mathcal{R}_j$ and $\mathcal{R}_k$. In addition, let us assume that the orbital motion of the bodies lies in the same plane and the spin vectors are perpendicular to it. From an arbitrary inertial frame, let ${\bf R}_i$, ${\bf R}_j$ and ${\bf R}_k$ be the respective position vectors of $m_i$, $m_j$ and $m_k$. We denote the relative position of $m_j$ to $m_i$ by ${\bf \Delta}_{ji}={\bf R}_j - {\bf R}_i$, while the angle subtended between the position of $m_j$ and the reference axis, as seen from $m_i$, is called $\varphi_{ji}$ (the same for $j=k$, see Figure \ref{3c}). 

Our goal is to study the spin dynamics of these three bodies under the effects of their mutual tidal perturbations. The tides will be modeled employing the creep theory \citep{FerrazMello2013,FerrazMello2015} in its closed form \citep{Folonier2018} and extended to the case of three interacting masses.

\subsection{Equilibrium figure}

\begin{figure*}[h!]
\centering
\includegraphics*[width=0.9\textwidth]{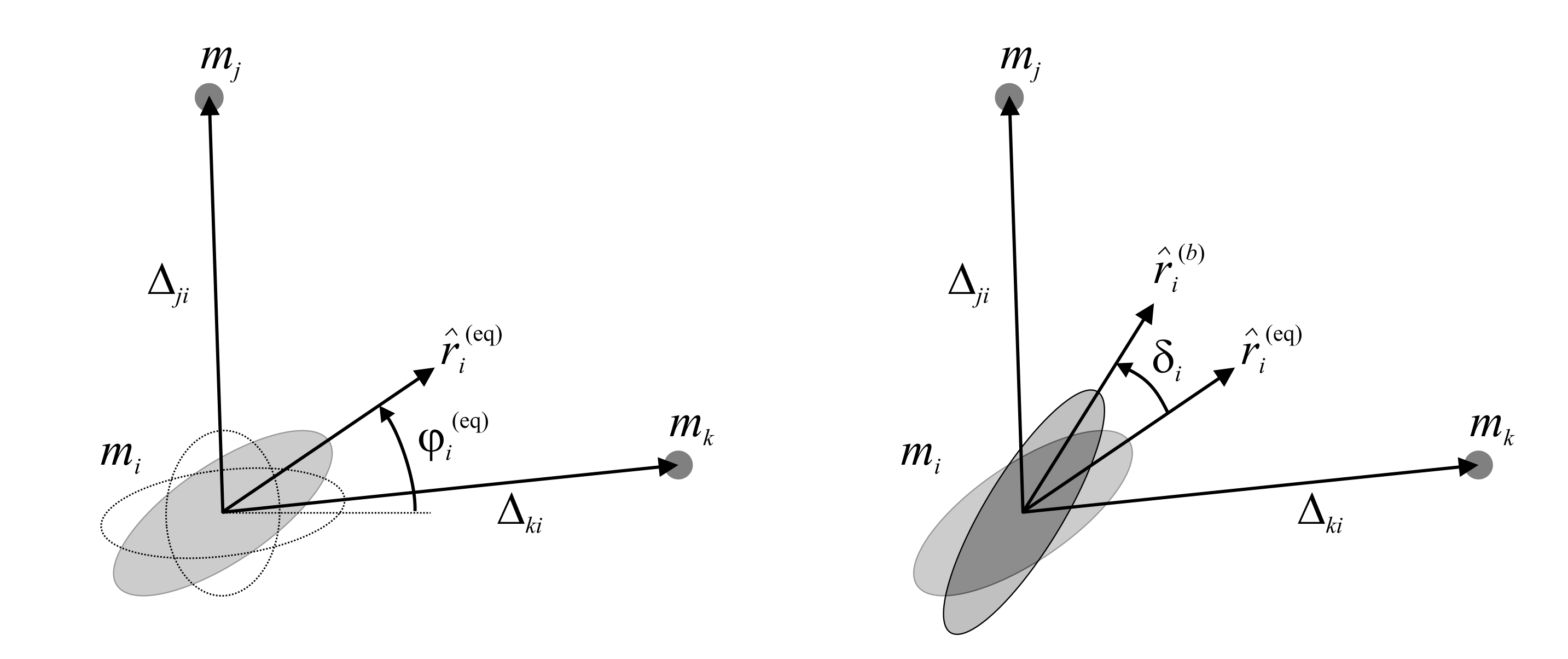}
\caption{{\it{Left: Equilibrium figure in the 3BP}}. Dashed-empty ellipsoids correspond to
the static tidal deformations generated by each of the perturbing masses, while the resulting equilibrium figure of $m_i$ (equation (\ref{eq:ronew})) is shown by a filled gray ellipsoid. The unit vector 
${\hat {\bf r}}^{(eq)}_i$ indicates the orientation of the equivalent equilibrium bulge. {\it{Right: Real figure in the 3BP}}. The dark grey filled ellipsoid represents the modelled real figure induced on $m_i$ when we take into account the internal viscosity. Note that it is displaced by an angle $\delta_i$ respect to the equilibrium figure and also has different flattenings.} 
\label{fig2}
\end{figure*}

As a consequence of the gravitational interaction on extended bodies, each mass $m_i$ undergoes a deformation which is given by the superposition of the tidal interaction with the two companions $m_j$ and $m_k$. In this work, we also consider the rotational flattening on each body due to its own spin, and assume that the resulting deformation due to these effects is small enough that a model developed up to the first order of the flattenings (like the one we will develop in this work) represents an accurate description of the problem.  

If we consider that each mass $m_i$ is a perfect fluid that instantly responds to any external stress, it will immediately adopt an equilibrium figure determined by the tides raised by the two companions (i.e. static tides) and its own rotation. The surface equation of this equilibrium figure seen from the center of mass, is determined by giving the distance $\rho_i$ of an arbitrary point on its surface with co-latitude angle $\theta_i$ and longitude angle $\varphi_i$. It can be expressed as the sum of the interactions between pairs as \citep{Tisserand1891,Folonier2015}
\begin{equation}
\begin{split}
\rho_{i}({\theta_i},{\varphi_i}) &=  \mathcal{R}_i \bigg[ 1 + \bigg( \epsilon_i^M +  \frac{\epsilon_{ij}^{\rho}}{2} + \frac{\epsilon_{ik}^{\rho}}{2} \bigg) \biggl( \frac{1}{3} - \cos^2{ \theta_i} \biggr) \\
&+ \frac{1}{2} \sin^2{\theta_i}  \biggl( \epsilon^{\rho}_{ij} \cos{(2{
\varphi_i} - 2 \varphi_{ji})} + \epsilon^{\rho}_{ik} \cos{(2{ \varphi_i} - 2 \varphi_{ki})} \biggr)\bigg] 
\end{split}
\label{eq:roi}
\end{equation}
where we have chosen the same definitions for the flattening as in \cite{Folonier2018}, given by
\begin{eqnarray}
\label{eij}
\epsilon_{ij}^{\rho} = \frac{15 m_j \mathcal{R}_i^3}{4 m_i \Delta^3_{ji}} \ \ &,& \ \
\epsilon_i^M = \frac{5 \mathcal{R}_i^3 \Omega_i^2}{4 m_i \mathcal{G}},  
\end{eqnarray}
with $\Omega_i$ the spin module of $m_i$ and $\mathcal{G}$ the gravitational constant.

We propose that the equilibrium figure expressed in equation (\ref{eq:roi}) corresponds to an equivalent triaxial flat ellipsoid with equatorial flattening $\epsilon_i^{\rho}$, polar flattening $\epsilon_i^{z}$ and equilibrium orientation defined by the unit vector ${\hat {\bf r}}^{(eq)}_i = (\cos{\varphi^{(eq)}_i},\sin{\varphi^{(eq)}_i})$ (see left panel of Figure \ref{fig2}). The surface equation of this equivalent equilibrium figure is simply given by  
\begin{equation}
\begin{split}
\rho_{i}({\theta_i},{\varphi_i}) 
&= \mathcal{R}_i \bigg[1 + \epsilon^{z}_i \bigg( \frac{1}{3} - \cos^2{\theta_i} \bigg) + \frac{\epsilon^{\rho}_{i}}{2}  \sin^2{\theta_i} \cos{(2\varphi_i-2\varphi_i^{eq})}  \biggr]
\end{split}
\label{eq:ronew}
\end{equation}

The equivalent flattenings and orientation can be obtained in terms of the two-body interacting parameters by equating equation (\ref{eq:ronew}) and equation (\ref{eq:roi}). This results in  
\begin{eqnarray}
\label{eq:formaeq}
\epsilon_i^{z} &=& \frac{1}{2}\bigg( \epsilon_{ij}^{\rho} + \epsilon_{ik}^{\rho}  \bigg) + \epsilon_{i}^{M} \nonumber \\
{(\epsilon_i^{\rho})}^2 &=& {(\epsilon_{ij}^{\rho})}^2 + {(\epsilon_{ik}^{\rho})}^2 + 2 \epsilon_{ij}^{\rho} \epsilon_{ik}^{\rho} \cos{(2\varphi_{ji}-2\varphi_{ki})}  \\
\tan{(2 \varphi_i^{eq})} &=& \frac{ \epsilon_{ij}^{\rho}\sin{(2\varphi_{ji})}+\epsilon_{ik}^{\rho}\sin{(2\varphi_{ki})} }{\epsilon_{ij}^{\rho}\cos{(2\varphi_{ji})}+\epsilon_{ik}^{\rho}\cos{(2\varphi_{ki})}}. \nonumber
\end{eqnarray}

\subsection{Creep equations}

Now, we assume the more realistic case in which each body has a certain viscosity that does not allow it to attain the equilibrium figure instantly. The real shape of $m_i$, measured by the surface distance $\zeta_{i}({ \theta_i},{\varphi_i})$, is also modelled as a triaxial flat ellipsoid but characterized by the equatorial flattening ${\mathcal E}_i^{\rho}$ and polar flattening ${\mathcal E}_i^{z}$. On the other hand, the orientation of the real figure is displaced respect to the equilibrium orientation by a lag $\delta_i$. Therefore, the real orientation unit vector is ${\hat {\bf r}}^{(b)}_i = (\cos{(\varphi^{(eq)}_i+\delta_i)}, \sin{(\varphi^{(eq)}_i+\delta_i)})$ (see right panel in Figure \ref{fig2}).
Thus, the surface equation of the real flat ellipsoid is then given by
\begin{equation}
\zeta_i(\theta_i,\varphi_i) =\mathcal{R}_i  
\bigg[ 1 +  {\mathcal E}_i^{z} \bigg( \frac{1}{3} - \cos^2{\theta_i} \bigg) +\frac{{\mathcal E}_i^{\rho}}{2}  \sin^2{\theta_i} \cos{(2 \varphi_i -2 (\varphi_{i}^{eq} +\delta_i))}  \bigg].
\label{eq9}
\end{equation}

Contrary to Darwin-based tidal models, in the creep tide theory we assume that $({\mathcal E}_i^{\rho}, {\mathcal E}_i^{z},\delta_i)$ are {\it{a-priori}} unknown quantities, whose values at each instant of time are obtained by solving the creep equation:
\begin{eqnarray}
\label{eq:creep}
\dot{\zeta}_i + \gamma_i \zeta_i = \gamma_i \rho_i ,
\end{eqnarray}
where $\gamma_i$ is the so-called relaxation factor 
of $m_i$ \citep{FerrazMello2013} which is inversely proportional to the body viscosity and will be assumed as constant in this work.

A set of three differential equations for the shape and orientation of the real figure of $m_i$ can be obtained by differentiating equation (\ref{eq9}) with respect to time, introducing its expression plus equations (\ref{eq:ronew}) and (\ref{eq9}) into equation (\ref{eq:creep}) and identifying terms of equal dependence with $(\theta_i,\varphi_i)$ (see \cite{Folonier2018}). Particularly, it is convenient to express the equatorial flattening and orientation in terms of regular variables as in \cite{Gomes2019} as

\begin{empheq}[box=\fbox]{align}
\label{eq:forma}
\dot{X_i} &= -\gamma_i X_i -2 Y_i (\Omega_i-\dot{\varphi_i}^{eq})+\gamma_i \epsilon_i^{\rho} \nonumber \\
\dot{Y_i} &= -\gamma_i Y_i +2 X_i (\Omega_i-\dot{\varphi_i}^{eq})\\
\dot{{\mathcal E}_i^{z}} &= \gamma_i ({\epsilon_i}^{z} - {{\mathcal E}_i}^{z}) \nonumber
\end{empheq}
where the regular variables are defined as 
\begin{eqnarray}
X_i = {\mathcal E}_i^{\rho} \cos(2\delta_i) \ \ \ \ &;& \ \ \ \ Y_i = {\mathcal E}_i^{\rho} \sin(2\delta_i).
\end{eqnarray}

By comparing equations (\ref{eq:forma}) with the shape and orientation evolution equations in the 2BP \citep[e.g.][]{Folonier2018,Gomes2019}, we note that in the 3BP the creep equations maintain the same form, which make the formalism applicable to any {\it N}-body problem. Thus, the complexity in the multi-body problem is only reduced to calculate the equilibrium figures. 

\subsection{Tidal forces and torques}

At any instant of time, given the rotation frequency $\Omega_i$ of the body $m_i$, it is possible to calculate its real shape and orientation by solving the differential equation system $(\ref{eq:forma})$. With these quantities, it is then possible to calculate the tidal forces and torques between each pair of bodies.

The tidal potential due to the deformation of $m_i$, induced on $m_j$ can be calculated as \cite[e.g.][]{Murray1999}
\begin{equation}
\begin{split}
U_{ji}({\bf{\Delta}}_{ji}) &= -\frac{{\cal G} (B_i-A_i)}{2\Delta_{ji}^3}\Big(3  ({{\bf{\hat{\Delta}}}_{ji}\cdot{\bf{\hat{r}}}_i^{b}})^2 - 1 \Big)-\frac{{\cal G} (C_i-B_i)}{2\Delta_{ji}^3},
\end{split}
\label{eq13}
\end{equation}
where $A_i,B_i,C_i$ are the principal moments of inertia of $m_i$, which can be related to the ellipsoidal flattenings (see Appendix A of \cite{Folonier2015}) as:
\begin{equation}
(B_i-A_i) \simeq \overline{\mathcal{C}}_i {\mathcal E}^{\rho}_i \hspace*{0.5cm} ; \hspace*{0.5cm} 
(C_i-B_i) \simeq \overline{\mathcal{C}}_i \biggl( {\mathcal E}^{z}_i - \frac{1}{2} {\mathcal E}^{\rho}_i 
\biggr) ,
\label{eq15}
\end{equation}
 where $\overline{\mathcal{C}}_i= (2/5) m_i \mathcal{R}_i^2$ is the principal moment of inertia of a homogeneous  spherical body of mass $m_i$ and radius $\mathcal{R}_i$.
 


The tidal force applied on $m_j$ due to the deformation of $m_i$ can be simply calculated as:
\begin{eqnarray}
\label{eq:fuerza}
{\bf{F}}_{ji} &=& -m_j \nabla_{{\bf \Delta}_{ji}}U_{ji} \nonumber \\
&=& \frac{3{\cal G}m_j \overline{\mathcal{C}}_i}{2\Delta_{ji}^5}{\mathcal E}^{\rho}_i\sin{(2(\varphi^{eq}_i+\delta_i)-2\varphi_{ji})}\ (\hat{\bf{k}}\times{\bf{\Delta}}_{ji}) \nonumber \\
&-&\frac{3{\cal G}m_j \overline{\mathcal{C}}_i}{2\Delta_{ji}^5} \Big(\frac{3}{2}{\mathcal E}^{\rho}_i\cos{(2(\varphi^{eq}_i+\delta_i)-2\varphi_{ji})}+{\mathcal E}^{z}_i\Big){\bf{\Delta}}_{ji},
\end{eqnarray}
where we have decomposed $\hat {\bf{r}}_i^{b}$ in terms of the radial unit vector $\hat{\bf{\Delta}}_{ji}$ and the azimutal unit vector $(\hat{\bf{k}} \times \hat{\bf{\Delta}}_{ji})$ ($\hat{\bf{k}}$ is the unit vector paralell to the spin vectors).

The torque induced on $m_j$ due to the deformation of $m_i$ can be simply calculated as 
\begin{equation}
\begin{split}
{\bf{T}}_{ji} &= {\bf{\Delta}}_{ji} \times {\bf{F}}_{ji}  \\
              &= \frac{3{\cal G}m_j \overline{\mathcal{C}}_i}{2\Delta_{ji}^3}{\mathcal E}^{\rho}_i\sin{(2(\varphi^{eq}_i+\delta_i)-2\varphi_{ji})}\ \hat{\bf{k}},
\end{split}
\label{eq:tji}
\end{equation}
while a completely analog expression can be found for the torque applied on $m_k$ due to the deformation of $m_i$ by exchanging the indexes.

Now, we assume that the reaction torque due to the deformation on $m_i$ is the only responsable in changing its own rotational velocity. This is a natural assumption that satisfies the condition that the variation of the rotation speed of $m_i$ should be  cancelled in the limit $\mathcal{R}_i \to 0$   \citep{Zoppetti2019}. Thus, the rotational evolution of $m_i$ is given by
\begin{eqnarray}
C_i \dot{{\bf{\Omega}}}_{i} + \dot{C}_i {{\bf{\Omega}}}_{i} &=& - \Big( {{\bf{T}}}_{ji} + {{\bf{T}}}_{ki}  \Big) \nonumber \\
&=&-\frac{3{\cal G} \overline{\mathcal{C}}_i{\mathcal E}^{\rho}_i}{2}\bigg[\frac{m_j}{\Delta_{ji}^3}\sin{(2(\varphi^{eq}_i+\delta_i)-2\varphi_{ji})} \nonumber \\
&+&\frac{m_k}{\Delta_{ki}^3}\sin{(2(\varphi^{eq}_i+\delta_i)-2\varphi_{ki})}\ \bigg]  \hat{\bf{k}}
\label{eq:torque}
\end{eqnarray}

Neglecting the term corresponding to the variation of the polar moment of inertia (i.e. $\dot{C}_i \sim 0$), using equations (\ref{eq:formaeq}) and after some algebra, it is possibly to find that the rotational evolution of the body $m_i$ is governed by 
\begin{empheq}[box=\fbox]{align}
\label{eq:spins}
\dot{\Omega}_{i} &= -\frac{2{\cal G} m_i}{5 \mathcal{R}_i^3} \epsilon_i^{(\rho)} Y_i 
\end{empheq}

The set of differential equations formed by equations
(\ref{eq:forma}) and (\ref{eq:spins}) described the dynamical evolution of the system on short timescales and will be target of this work. 

\section{Numerical simulations}
\label{sec:nume}

Circumbinary planetary systems are particularly interesting scenarios where we can study the effects of tides on the dynamical evolution, in the framework of the 3BP. In this context, the stellar tidal evolution is expected to be very little affected by the presence of the CB planet and mainly dominated by the 2-body interaction between the binary components \citep{Correia2016,Zoppetti2019}. 
Since the rotational evolution due to tides in a binary system has been extensively studied \cite[e.g.][]{Mignard1979,Hut1981,Efroimsky2012}, in this paper we will focus on the evolution of the planet.

Using a CTL model based on the tidal forces and torques of \cite{Mignard1979}, in \cite{Zoppetti2019,Zoppetti2020} we found that the characteristic timescales of the rotational evolution of any planet in a {{\it Kepler}}-like circumbinary system are much faster than the characteristic timescales of the orbital evolution of the system. As a result, in this framework, it is expected that while the shape, orientation and rotation of each body reach stationary solution, only periodic oscillations be observed in the orbital elements of the bodies, except for the fast angles. 
For this reason, in this work we take advantage of this adiabatic nature of the problem and solve the set of differential equation associated with the evolution of the shape of each body, its orientation and its rotation (equations (\ref{eq:forma}) and (\ref{eq:spins})), assuming fixed values for all the orbital elements except the true anomalies. The orbital evolution of the bodies on much longer timescales will be addressed in a forthcoming work.

With this motivation, we begin first by analyzing the results of the numerical integrations of the equations (\ref{eq:forma}) and (\ref{eq:spins}) in a 3BP consisting of a single planet around a binary star. To compare with the results of our previous works \citep{Zoppetti2018,Zoppetti2019}, we choose a Kepler-38 like system as a test case.  Nominal values for system parameters and initial orbital elements are detailed in Table 1. Stellar masses and radii were taken close to the observed by \cite{Orosz2012}, while the value of $m_2$ was estimated from a semi-empirical mass-radius fit \citep{Mills2017}.
\begin{table}
\centering
\caption{Initial conditions of our reference numerical simulation, representing a Kepler-38 like system \protect\citep{Orosz2012}. Orbital elements are given in a Jacobi reference frame. The parameters highlighted with an asterisk were varied in different simulations as is indicated in the text.}
\label{tab1}
\begin{tabular}{lccc} 
\hline
body & $m_0$ & $m_1$ & $m_2$ \\
\hline
mass                       & $0.95 \, m_\odot$ & $0.25 \, m_\odot$(*)   & $10 \, m_\oplus$   \\
radius                     & $0.84 \, \mathcal{R}_\odot$  & $0.27 \, \mathcal{R}_\odot$   & $4.35 \, \mathcal{R}_\oplus$ \\
$a_i$ [AU]                 &                    & $0.15$               & $0.45$(*)             \\
$e_i$                      &                    & $0.15$               & $0.05$(*)          \\
$\Omega_i$                 & $5 \, n_1$     & $5 \, n_1$       & $5 \, n_2$     \\
\hline
\end{tabular}
\end{table}
\begin{figure*}[h!]
\centering
\includegraphics*[width=0.8\textwidth]{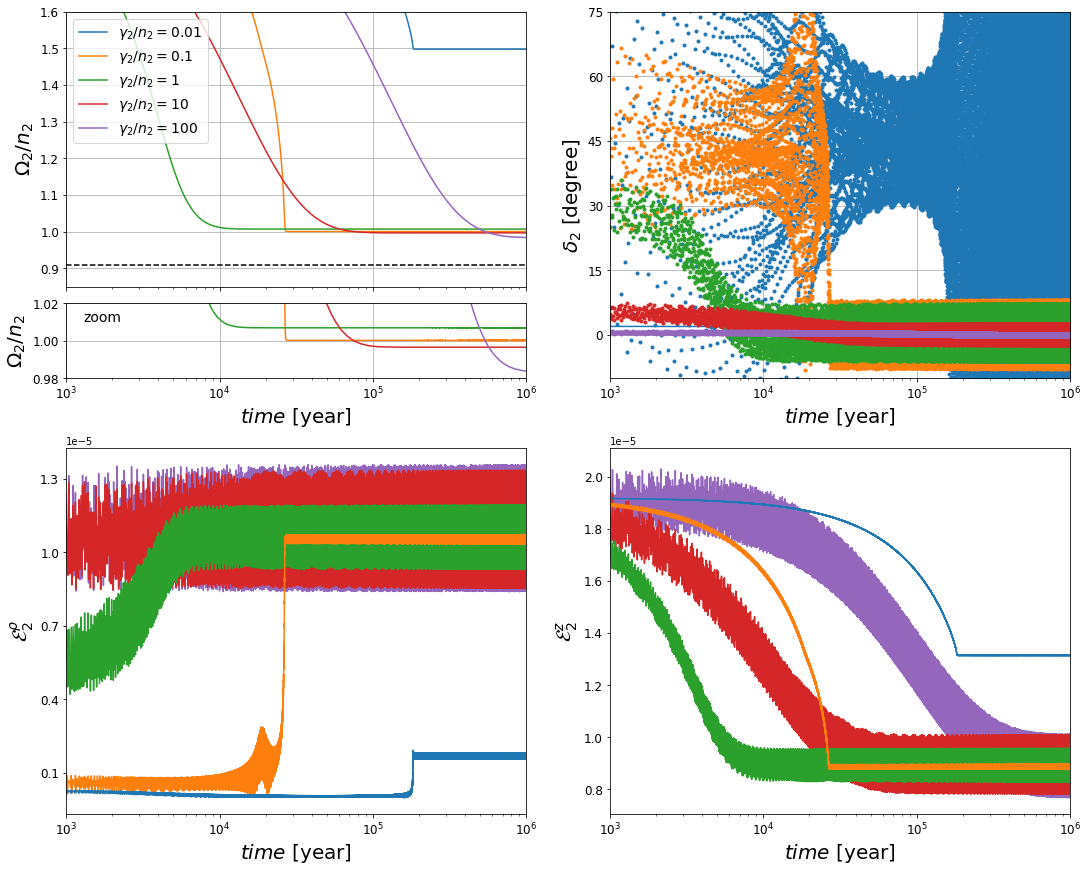}
\caption{Rotational evolution due to tides in a Kepler-38 like CB planet (see Table \ref{tab1}), considering different values of planetary relaxation factor $\gamma_2$ with different colors (see legend in the upper left panel). The evolution was calculated by direct numerical integration of equations (\ref{eq:forma}) and (\ref{eq:spins}). The evolution of the planetary spin rate $\Omega_2$ is shown in upper left panel (zoomed around $\Omega_2 \sim n_2$ in a secondary plot), the lag-angle $\delta_2$ in upper right panel, the equatorial flattening $\mathcal{E}^\rho_2$ in the lower left panel and the polar flattening $\mathcal{E}^z_2$ in the lower right panel. The dashed black line in the upper left panel shows the stationary solution predicted by \cite{Zoppetti2019} with a CTL-based model.}
\label{fig:caidas}
\end{figure*}

The first obstacle we encountered when doing the numerical integration was the virtual impossibility of solving the equations system numerically for bodies with large relaxation factors ($\gamma_2 >> n_1, n_2$), typically associated with bodies with a large gaseous component \citep{FerrazMello2013}. 
This limitation comes from the fact that, for large $\gamma$-values, the shape and orientation of the deformed body quickly adapt to the change in position of the deforming bodies, so the integrator must greatly reduce the time-step to pursue it and, finally, integration over millions of years timescales becomes computationally impracticable \citep[e.g.][]{Lambert1991}. However, as we will see, the rotational evolution timescales can be extrapolated from the simulations of more viscous bodies in the same regime, and the synchronous attractor from the analytical solution calculated in Section \ref{sec:ana}. 

In Figure \ref{fig:caidas} we show the rotational evolution of a CB planet from a Jacobi reference frame, obtained by direct numerical integration of equations (\ref{eq:forma}) and (\ref{eq:spins}). 
All initial conditions and system parameters were taken equal to the nominal values summarized in Table \ref{tab1}, while the angles were set equal to zero. As we mentioned above, the orbital elements are considered fixed, with the exception of the mean anomalies which vary linearly with the time. 

Many interesting features can be noticed in Figure \ref{fig:caidas}. We begin first by identifying, as in the 2BP, two different regimes whose boundary value is close to the main dynamical frequency of the problem $n_2$: the "stiff regime" for $\gamma_2<n_2$ and the "gaseous regime" for $\gamma_2>n_2$. In the stiff regime, the rotational evolution acquires its stationary value faster the higher the relaxation factor $\gamma_2$, while the opposite occurs in the gaseous limit: if we decrease the planetary viscosity (increase $\gamma_2$) the attractor is reached on longer timescales. This bimodal character of the energy dissipation has already been observed previously in the 2BP \citep{FerrazMello2013,Correia2014}: for relaxation parameters far from the maximum, the energy dissipation decreases proportional to $\gamma_2$ in the gaseous regime and decreases inversely proportional to $\gamma_2$ in the stiff bodies limit. 

In the stiff regime, it is possible to notice from the upper left panel of Figure \ref{fig:caidas} that for bodies with very small relaxation factors $\gamma_2/n_2 \sim 0.01$, ( i.e. $\gamma_2 \simeq 7 \times 10^{-9}$ Hz), the planetary spin captures may occur in an attractor different than 1:1 spin-orbit resonance, in our testing case 3:2 with an associated circularization of the lag-angle $\delta_2$. This result is expected from the 2BP experience, where for very low $\gamma_2$-values, the probability of capture in this type of resonance increases considerably \citep{Correia2014,Gomes2019}. However, according to our numerical simulations, captures in resonances different than the 1:1 seems to occur for bodies with satellite-like relaxation factors like the moon, which are expected to be approximately two orders of magnitude lower than those expected for rocky earth-type planets \citep{FerrazMello2013}, or for bodies with high initial eccentricities, a case that we will not consider in this work.
This particular stationary state also has a counterpart in the flattenings $\mathcal{E}^\rho_2$ and $\mathcal{E}^z_2$, which oscillate around values  significantly different than in the case of the capture in the 1:1 spin-orbit resonance (see lower panels in Figure \ref{fig:caidas}).

The oscillation of the lag-angle around $\delta_2 \sim 45^\circ$ of free rotating bodies in the stiff regime can be observed in the upper right panel of Figure \ref{fig:caidas} (blue and orange dots). This result also has been previously reported in the framework of the 2BP \citep{Gomes2019} and is mainly due to the lack of the elastic tide component inside the creep model \citep{FerrazMello2015b}. Finally, we notice that in this limit, the oscillation of $\delta_2$ (and also of $\Omega_2$, although it can not be distinguished in the upper left panels) around the stationary values are of much greater amplitudes respect to those in the gaseous regime, while the completely opposite situation occurs for $\mathcal{E}^\rho_2$ and $\mathcal{E}^z_2$. As noticed by \cite{Folonier2018} in the 2BP, this periodic oscillations of large amplitudes must be taken into account when constructing an analytical approach (see Section \ref{sec:ana}).
 
Regarding the gaseous regime, despite considering bodies of different viscosities (i.e. relaxation factors $\gamma_2$), we observe similar behaviors: stationary spins are located very close to $n_2$ and the lag-angles $\delta_2$ oscillates close to zero, with low oscillation amplitudes. If we omit the case of $\gamma_2/n_2 = 1$ which is a behavior change value, we can observe in the secondary plot of the upper left panel that the equilibrium spins become smaller as we increase $\gamma_2$ and turn out to be sub-synchronous for these parameter system in the gaseous regime (see red and magenta curves).
However, despite the fact that the rotational equilibrium state resulting from the numerical integrations appears to be sub-synchronous in the gaseous limit, this solution is far from the one we found in \cite{Zoppetti2019} (dashed black line in the upper left panel). The details of the 1:1 spin-orbit mean solution will be described in Section \ref{sec:ana}, while the reasons for the discrepancy with \cite{Zoppetti2019} will be discussed in Section \ref{sec:Mignard}.

\section{An analytical solution for the mean rotational stationary state}
\label{sec:ana}

As we have discussed in the previous Section \ref{sec:nume}, according to our numerical simulations, the rotational evolution due to tides of a CB planet in a low-eccentric orbit with $\gamma_2 \gtrsim 0.1 n_2$ (approximately the expected range for a planet-like body, according to \cite{FerrazMello2013}) tends to be captured in the 1:1 spin-orbit resonance. 

The timescales for the capture depend on the planetary viscosities. While for bodies in the stiff regime are expected to be $\lesssim$ Myrs, in the gaseous regime can be of the order of some Gyrs, considering $\gamma_2$-value similar to the one estimated for Neptune \citep{FerrazMello2013} and extrapolating the behavior observed in the Figure \ref{fig:caidas} by assuming a linear dependence of the decay times with the relaxation factors. Although for giant planets these timescales may be even larger than the age of the systems, the enormous uncertainty that exists about the internal structure of the exoplanets, and in particular the CB {\it{Kepler}} planets, prohibits a reliable estimate of the planetary relaxation factor. In addition to this, the central binary could have undergone an important change in both the physical parameters and the orbital elements, which may have accelerated the tidal evolution processes \citep{Zoppetti2018}. In summary, according to our model, a CB planet in a Kepler-38 like system with a large rocky component is expected to have reached its stationary spin and this is also possible for the case of gaseous CB planets with small relaxation factors $\gamma_2$ .

From a practical point of view, having an analytical approach for the rotational stationary state will allow us, when studying long-term orbital evolution of the system, to avoid direct numerical integration of the equations (\ref{eq:forma}) and (\ref{eq:spins}), which are particularly difficult to solve for bodies with large relaxation factors (see Section \ref{sec:nume}) and, instead, to work with a semi-analytical approach in which we introduce a recipe for the equilibrium shape, orientation and spin. 

With these motivations, in this Section we develop an analytical approximation for the equations system (\ref{eq:forma}) and (\ref{eq:spins}) around the pseudo-synchronous stationary state, in this particular scenario where there are two central bodies (both stellar components of the binary) exerting tides onto a planet. 
Mathematically speaking, our goal is to find the stationary solution for the system formed by $(X_2,Y_2,{\mathcal E}^{z}_2)$ and $\Omega_2$, employing a power series truncated in $\alpha=a_1/a_2$, $e_1$ and $e_2$.

To do this, we first adopt a Jacobi reference frame for the coordinates and velocities of the bodies, where the state vector of $m_1$ is defined as $m_0$-centric, while the state vector of $m_2$ is measured with respect to the barycenter of $m_0$ and $m_1$. We note that these coordinates and velocities are required to compute the forcing terms given by $\dot{\varphi_i}^{eq}$, $\epsilon_i^{\rho}$ and ${\epsilon_i}^{z}$, in our differential equation system. 

Regarding the resolution method, we choose the same path as the one taken in Appendix 2 of \cite{Folonier2018}. It consists in proposing a particular stationary solution, inspired in the form (i.e. dependence on fundamental frequencies) of the elliptical expansions of the forced terms $\epsilon^{\rho}_i$ and $\dot{\varphi_i}^{eq}$. In this case, we propose a solution of the form 
\begin{eqnarray}
\label{eq:solugral}
\Omega_2 = \sum_{\vec{\ell}} \{\Omega_2\}_{\vec{\ell}}  \cos{(l_1 M_1 + l_2 M_2 + l_3 \varpi_1 + l_4 \varpi_2-\Phi_{\vec{\ell},\Omega_2})} \nonumber \\ 
X_2 = \sum_{\vec{\ell}} \{X_2\}_{\vec{\ell}}  \cos{(l_1 M_1 + l_2 M_2 + l_3 \varpi_1 + l_4 \varpi_2-\Phi_{\vec{\ell},X_2})}\nonumber \\
Y_2 = \sum_{\vec{\ell}} \{Y_2\}_{\vec{\ell}}  \cos{(l_1 M_1 + l_2 M_2 + l_3 \varpi_1 + l_4 \varpi_2-\Phi_{\vec{\ell},Y_2})} \\
\mathcal{E}^z_2 =  \sum_{\vec{\ell}} \{\mathcal{E}^z_2\}_{\vec{\ell}}  \cos{(l_1 M_1 + l_2 M_2 + l_3 \varpi_1 + l_4 \varpi_2-\Phi_{\vec{\ell},\mathcal{E}^z_2})}, \nonumber 
\end{eqnarray}
where $M_1$ and $\varpi_1$ are the mean anomaly and pericenter longitude of the secondary star, while $M_2$ and $\varpi_2$ are those corresponding to the planet. The amplitudes $\{w\}_{\vec{\ell}}$ and the constant phases $\Phi_{\vec{\ell},x}$ depend on the subscripts $\vec{\ell}=(l_1,l_2,l_3,l_4)$ and each variable $w=\Omega_2,X_2,Y_2$ or $\mathcal{E}^z_2$.
This particular solution and their derivatives are then introduced into the system (\ref{eq:forma}) and (\ref{eq:spins}), and the coefficients can be calculated by equating terms with same trigonometric argument and neglecting terms of higher-order. The terms in the system (\ref{eq:solugral}) with $l_1=l_2=0$ correspond to the mean stationary solution and will be given in what follows. 

We choose to express the stationary solution of the rotational evolution written in terms of the mean planetary spin $\langle \Omega_2 \rangle$, mean planetary lag-angle $\langle \delta_2 \rangle$, mean planetary equatorial flattening $\langle {\mathcal E}^\rho_2 \rangle$ and
mean planetary polar flattening $\langle {\mathcal E}^z_2 \rangle$.
Up to 3rd-order in $\alpha=a_1/a_2$ and to 2nd-order in the eccentricities $e_1$ and $e_2$, including also the contribution of the eccentricities independent 4th-order and 6th-order terms, the solution is explicitly given by
\begin{eqnarray}
\label{eq:ome_st}
 \langle \Omega_2 \rangle &=& n_2 + 6\frac{\gamma_2^2 n_2}{\gamma_2^2+n_2^2}  e_2^2 
+ 5 \frac{\gamma_2^2n_2}{\gamma_2^2+n_2^2} e_2^2  \mathcal{M}_2 \alpha^2  \\
&-&  \frac{325}{16} \frac{\gamma_2^2n_2}{\gamma_2^2+n_2^2}  e_1 e_2 \cos(\Delta \varpi)  \mathcal{M}_3 \alpha^3  \nonumber \\
&-&  \frac{19}{2} \frac{\gamma_2^2(2n_1-2n_2)}{\gamma_2^2+{(2 n_1-2n_2)}^2}  \mathcal{M}_2^2 \alpha^4 \nonumber + \langle \Omega_2 \rangle_{o6}
\end{eqnarray}
\begin{eqnarray}
\label{eq:delta_st}
\langle \delta_2 \rangle &=&  6\frac{\gamma_2 n_2}{\gamma_2^2+n_2^2}  e_2^2 + 5 \frac{\gamma_2 n_2}{\gamma_2^2+n_2^2} e_2^2  \mathcal{M}_2 \alpha^2    \\
&-& \bigg[\frac{325}{16}\gamma_2\cos(\Delta \varpi)  +\frac{175}{32} n_2\sin(\Delta \varpi) \bigg]\frac{n_2}{\gamma_2^2+n_2^2} e_1 e_2 \mathcal{M}_3\alpha^3  \nonumber \\
&-&  \frac{19}{2}  \frac{\gamma_2 (2n_1-2n_2)}{\gamma_2^2+{(2 n_1-2n_2)}^2} \mathcal{M}_2^2 \alpha^4 + \langle \delta_2 \rangle_{o6} \nonumber 
\end{eqnarray}
\begin{eqnarray} 
\label{eq:er_st}
\frac{\langle {\mathcal E}^\rho_2 \rangle}{\overline{\epsilon}^\rho_2} &=&  1 + \bigg(\frac{3}{2} - 4 \frac{n_2^2}{\gamma_2^2+n_2^2} \bigg) e_2^2  \\ &+& \bigg[ \frac{5}{4}+\frac{15}{8}e_1^2 + \bigg( \frac{25}{4} - 5 \frac{n_2^2}{\gamma_2^2+n_2^2} \bigg) e_2^2 \bigg]  \mathcal{M}_2 \alpha^2  \nonumber \\
&-& \frac{25}{8} \frac{(3n_2^2+5\gamma_2^2)}{\gamma_2^2+n_2^2}  e_1 e_2 \cos(\Delta \varpi) \mathcal{M}_3 \alpha^3  \nonumber\\
&+& \bigg[ \frac{105}{64} \mathcal{M}_4 +  \frac{4\gamma_2^2\mathcal{M}_2^2}{\gamma_2^2 + {(2n_1-2n_2)}^2}  \bigg] \alpha^4 + \frac{{\langle \mathcal E}^\rho_2 \rangle_{o6}}{\overline{\epsilon}^\rho_2} \nonumber  
\end{eqnarray}
\begin{eqnarray}
\label{eq:ez_st}
\frac{\langle {\mathcal E}^z_2 \rangle}{\overline{\epsilon}^M_2} &=&  \bigg( 1 + 12\frac{\gamma_2^2}{\gamma_2^2+n_2^2} e_2^2 + 10 \frac{\gamma_2^2}{\gamma_2^2+n_2^2} e_2^2 \mathcal{M}_2 \alpha^2  \\
&-& \frac{325}{8} \frac{\gamma_2^2}{\gamma_2^2+n_2^2}  e_1 e_2 \cos(\Delta \varpi)  \mathcal{M}_3 \alpha^3  \nonumber \\
&-& 19\frac{{(2n_1-2n_2)}\gamma_2^2}{n_2(\gamma_2^2 + {(2n_1-2n_2)}^2)} \mathcal{M}_2^2\alpha^4 \bigg) \nonumber \\
&+& \frac{\overline{\epsilon}^\rho_2}{\overline{\epsilon}^M_2} \ \bigg( \frac{1}{2} + \frac{3}{4}e_2^2 \nonumber + \bigg[ \frac{9}{8}+\frac{27}{16}e_1^2 +\frac{45}{8}e_2^2 \bigg]  \mathcal{M}_2 \alpha^2  \\
&-& \frac{225}{16}  e_1 e_2 \cos(\Delta \varpi) \mathcal{M}_3 \alpha^3 + \frac{225}{128} \mathcal{M}_4  \alpha^4 \bigg) + \frac{{\langle \mathcal E}^z_2 \rangle_{o6}}{\overline{\epsilon}^M_2} \nonumber
\end{eqnarray}
where $\langle \Omega_2 \rangle_{o6}$, $\langle \delta_2 \rangle_{o6}$, 
${\langle \mathcal E}^\rho_2 \rangle_{o6}$ and 
${\langle \mathcal E}^z_2 \rangle_{o6}$
are the terms independent on the eccentricities, expanded up to 6th-order in $\alpha$ and listed in the Appendix. We mention that the 5th-order terms are eccentricities dependent and, thus, were excluded in this approximation. 
The mass factors are defined in terms of the binary mass ratio $\beta = m_1/m_0$ as
\begin{eqnarray}
\mathcal{M}_2 = \frac{\beta}{{(1+\beta)}^2} \ , \ \mathcal{M}_3 = \frac{\beta-\beta^2}{{(1+\beta)}^3} \ , \ \mathcal{M}_4 = \frac{\beta-\beta^2+\beta^3}{{(1+\beta)}^4} 
\end{eqnarray}
and the equilibrium mean flattenings are given by
\begin{eqnarray}
\overline{\epsilon}^\rho_2 = \frac{15 (m_0+m_1) \mathcal{R}_2^3}{4 m_2 a^3_2} \ \ \ &\mathrm{and}& \ \ \ \overline{\epsilon}^M_2 = \frac{5 \mathcal{R}_2^3 n_2^2}{4 m_2 \mathcal{G}}. 
\end{eqnarray}
\begin{figure*}[h!]
\centering
\includegraphics*[width=0.75\textwidth]{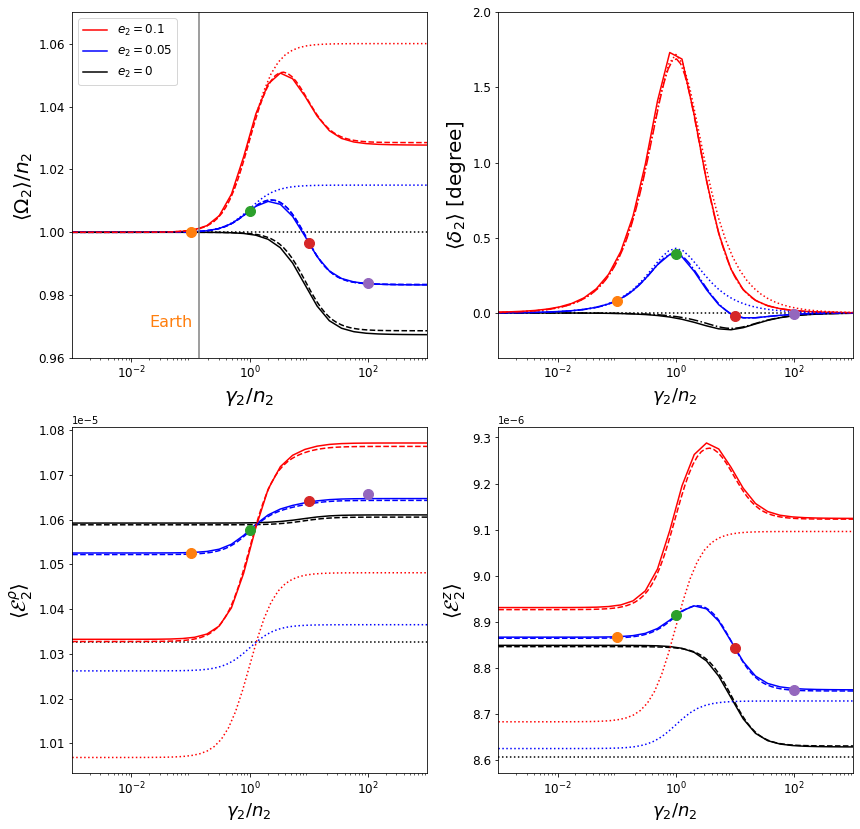}
\caption{Mean rotational speed $\langle \Omega_2 \rangle$ (upper left panel), mean lag-angle $\langle \delta_2 \rangle$ (upper right panel), mean equatorial flattening $\langle {\mathcal E}^\rho_2 \rangle$ (lower left panel) and mean polar flattening $\langle {\mathcal E}^z_2 \rangle$ (lower right panel) as a function of the planetary relaxation factor normalized by the mean motion frequency $\gamma_2/n_2$. The initial conditions for the system were taken from Table \ref{tab1}, but considering different planetary eccentricities shown with different color curves, as indicated in the upper left panel. Different types of curves represent different methods for obtaining the mean values: full curves represent the numerical results, dashed curves represent our 6th-order fit and dotted curves the results of considering the 2BP with a central mass equal to the sum of the binary components. Vertical grey line in the upper left panel represents the position in the diagram of a solid Earth \citep{FerrazMello2013} and big color dots represent the final mean value obtained with different $\gamma_2$-values, from simulations of Figure \ref{fig:caidas} captured in the 1:1 spin-orbit resonance: $\gamma_2/n_2 = 0.1$ (orange), $\gamma_2/n_2 = 1$ (green), $\gamma_2/n_2 = 10$ (red) and $\gamma_2/n_2 = 100$ (magenta). }
\label{fig:estacionario}
\end{figure*}
\begin{figure*}[h!]
\centering
\includegraphics*[width=0.755\textwidth]{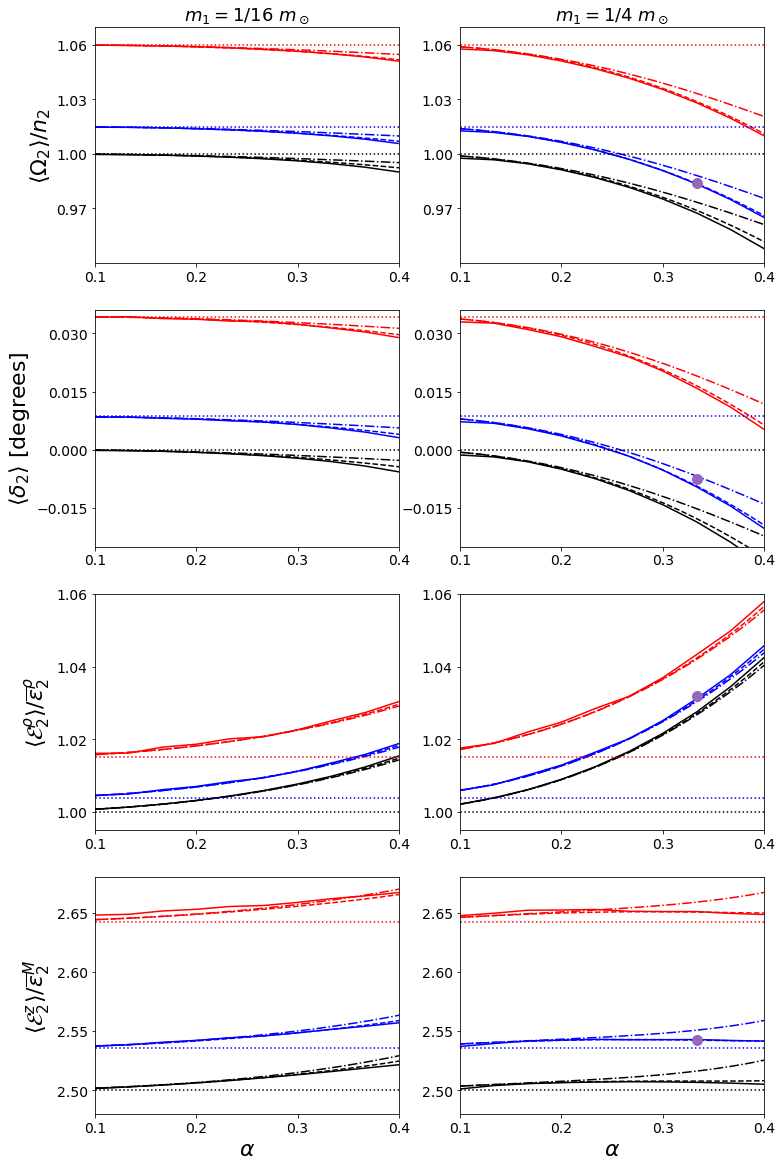}
\caption{Mean rotational speed $\langle \Omega_2 \rangle$ (first row panels), mean lag-angle $\langle \delta_2 \rangle$ (second row panels), mean equatorial flattening $\langle {\mathcal E}^\rho_2 \rangle$ (third row panels) and mean polar flattening $\langle {\mathcal E}^z_2 \rangle$ (last row panels) as a function of the semimajor axes ratio $\alpha = a_1/a_2$, for the case $\gamma_2 = 100 n_2$. The initial conditions for the system were taken from Table \ref{tab1}, but considering different planetary eccentricities shown with different color curves (same color convention as in Figure \ref{fig:estacionario}) and different secondary mass: $m_1 = 1/16 m_\odot$ (first column) and $m_1 = 1/4 m_\odot$ (second column).  Different types of curves represent different methods for obtaining the mean values with the same convention than in Figure \ref{fig:estacionario} but including the 4th-order fit with dotted-dashed curves. The magenta dots in the right panels represent the final mean value obtained for this relaxation factor from the simulations of Figure \ref{fig:caidas}.}
\label{fig:alfas}
\end{figure*}
\begin{figure*}[h!]
\centering
\includegraphics*[width=0.8\textwidth]{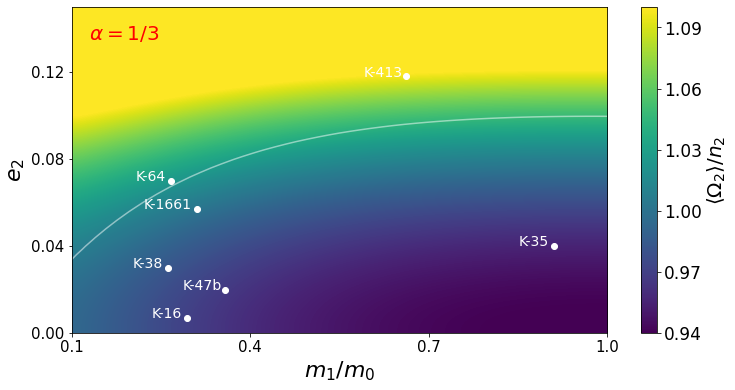}
\caption{Stationary spin map predicted by our analytical model (equation (\ref{eq:ome_st})) for a CB planet in the gaseous regime ($\gamma_2 >> n_2$) and in an orbit with $\alpha=a_1/a_2=1/3$, as a function of its eccentricity $e_2$ and the binary mass ratio $m_1/m_0$. The white curve indicate the perfectly synchronous state $\langle \Omega_2 \rangle / n_2 = 1$. The white dots represent the position in this diagram of the {\it{Kepler}} CB planets with expected large gaseous component located close to the binary.  }
\label{fig:keplers}
\end{figure*}

We first note from expressions (\ref{eq:ome_st}), (\ref{eq:delta_st}), (\ref{eq:er_st}) and (\ref{eq:ez_st}) that the mean solutions depend on the mean orbital configuration mainly through $e_2$ up to low-order in $\alpha$, but also  through $e_1$ and $\Delta\varpi$ up to higher-orders. The dependence of the expressions with the masses $m_0$ and $m_1$, and the semimajor axes ratio $\alpha$ are given in such a way that when $m_1 \to 0$ (i.e. a secondary star of very low mass) or $\alpha \to 0$ (i.e. a planet far away from the binary), we recover the 2BP expressions of \cite{Folonier2018}.

On the other hand, according to preliminary numerical simulations, the mean rotational solutions only exhibit a very weak dependence with $e_1$. This fact is also expected if we notice that, in the mean analytical expression, the functional dependence with this parameter only appears up to 3rd-order in $\alpha$. 

Another interesting feature of the mean solutions is the strong correlation between the stationary spin $\langle \Omega_2 \rangle$ with the lag-angle $\langle \delta_2 \rangle$: if we neglect the dependence on $\Delta \varpi$ (for example, for aligned and non-evolving pericenters or for the circulating case), these quantities are simply related as $\langle \Omega_2 \rangle - n_2= \langle \delta_2 \rangle/\gamma_2$. As a result, a sub or super-synchronous mean stationary rotation speed is directly associated with a delay or advance of the tidal bulge. 

The rheology of the planet is taken into account through the ratio mainly between the relaxation factor $\gamma_2$ and the main frequency of the problem $n_2$. However, up to higher $\alpha$-order, another frequencies such as $2 n_1 - 2 n_2$ starts become important in the solutions. For this last reason, as we comment in Section \ref{sec:nume}, the exact position of the frontier between the stiff regime and the gaseous regime is expected to be close to $\gamma_2 = n_2$, but not exactly at this point. 

In Figure \ref{fig:estacionario}, we compare the mean values of $\langle \Omega_2 \rangle$, $\langle \delta_2 \rangle$, $\langle {\mathcal E}^\rho_2 \rangle$ and $\langle {\mathcal E}^z_2 \rangle$ obtained with different methods, as a function of the normalized relaxation factor $\gamma_2/n_2$. The full curves indicate the results of a numerical integration from an initial condition far from the equilibrium which is allowed to evolve towards the stationary state and then is filtered over the fast angles. The dashed curves represent the results obtained respectively with equations (\ref{eq:ome_st}) , (\ref{eq:delta_st}), (\ref{eq:er_st}) and (\ref{eq:ez_st}), while the dotted curves represent the simplified 2BP case in which the central binary is replaced by a unique body with mass equal to the sum of the binary components. Big color dots represent the final mean value obtained with different $\gamma_2$-values in simulations of Section \ref{sec:nume}, which result in a capture in the 1:1 spin-orbit resonance (see Figure \ref{fig:caidas}): the case of $\gamma_2/n_2 = 0.1$ (in orange), $\gamma_2/n_2 = 1$ (in green), $\gamma_2/n_2 = 10$ (in red) and $\gamma_2/n_2 = 100$ (in magenta). We note that, in this last case, the magenta dots are not perfectly located on the blue full curves due to this final condition, taken from simulation of Figure \ref{fig:caidas}, is almost but still not perfectly in the stationary state.


Figure \ref{fig:estacionario} allows us to observe the existence of the two different regimes commented in Section \ref{sec:nume}: the stiff regime for $\gamma_2 < n_2$ and the gaseous regime for $\gamma_2 > n_2$, where far enough away from the transition value $\sim n_2$, the rotational equilibrium solution becomes independent of the planetary relaxation factor $\gamma_2$. In particular, in the case of the mean stationary speed $\langle \Omega_2 \rangle$ (upper left panel), planets in the stiff regime are expected to asymptotically synchronize their spins with the mean motion, independently of their orbital parameters, body masses and viscosities. This prediction is completely analog to that expected with the 2BP \citep{Folonier2018}. However, a much greater wealth of behavior occurs in the gaseous regime: the stationary solutions do depend on the orbital parameters and binary component masses. In this regime, unlike what is expected in the 2BP (dotted curves), the presence of an inner secondary star leads to lower stationary spins which can be sub-synchronous, in the low planetary eccentric limit. 

With respect to the accuracy of our analytical model in fitting the filtered numerical integrations, we can observe that is not only able to predict the sub-synchronous solution of the spins, but also gives a very precise estimation of the equilibrium solution in the complete range of $\gamma_2$-values. On the other hand, the 2BP simplification can also give a good prediction, but the accuracy of the approximation is strongly dependent on the initial condition and the relaxation factor range. 

In Figure \ref{fig:alfas}, we evaluate with more detail the accuracy of our mean stationary solution given by expressions (\ref{eq:ome_st}), (\ref{eq:delta_st}), (\ref{eq:er_st}) and (\ref{eq:ez_st}), as a function of the small parameters $\alpha=a_1/a_2$ and $e_2$, for the gaseous case $\gamma_2 = 100 n_2$. Different $\alpha$-values were obtained by varying the semi-major axis of the planet $a_2$. Although it is not a small parameter of our model, we also explore a different secondary mass $m_1$. 

We observe from Figure \ref{fig:alfas} that our 6th-order analytical fit is a very precise estimation of the filtered numerical integrations for the rotational evolution of a CB planet in a Kepler 38-like system. This occurs even for high $\alpha$-values such as $\alpha \sim 0.4$, where the planet is probably inside the instability limit imposed by the presence of an internal secondary disturber \citep[e.g.][]{Holman1999}. A good approximation is also obtained with the 4th-order fit. However, because most Kepler CB planets are located very close to binary (typically $\alpha \sim 0.3$), a higher order approach is required to model more accurately its mean rotational behavior.

As expected, for the standard case ($m_1 = 1/4 m_{\odot}$, right column) the 2BP simplification only is a good fit of the numerical simulation for low $\alpha$-values, where the tidal effects are probably of little importance. Curiously, the case of the mean polar flattening $\langle {\mathcal E}^z_2 \rangle$ presents an exception: the numerical stationary solutions show little dependence with $\alpha$ and, while the 6th-order fit is the best fit in the complete range, the 2BP simplification is also a good approximation, even better than the 4th-order fit for planetary positions very close to the binary. This may seem less strange if we remember the expression of the equilibrium polar flattening $\epsilon_i^{z}$ given in equation \ref{eq:formaeq}. Due to the definition of the flattenings chosen in this study (see \cite{Folonier2018}), $\epsilon_2^{z}$ contains terms that depend on the equatorial flattenings $\epsilon_{2,0}^{\rho}$ and $\epsilon_{2,1}^{\rho}$ which increase for higher $\alpha$-values, in addition to the classical Maclaurin flattening $\epsilon_{2}^{M}$ which decreases as we move closer to the binary, since the stationary planetary spin does. Thus, we expect a similar behavior in the real polar flattening $\mathcal{E}_2^{z}$ that,  due to a potential balance of this two contributions, may result in a weak dependence with $\alpha$.

The case of $m_1 = 1/16 m_\odot$ (left panels of Figure \ref{fig:alfas}) is considered as an illustrative example that exhibits the strong dependence of the solution with the secondary mass: for a low-mass secondary companion, the 2BP simplification begins to be a good approximation, specially for CB planets in wide orbits. 

As an application example of our model predictions, in Figure \ref{fig:keplers} we show a stationary spin map constructed with equation (\ref{eq:ome_st}), for a CB planet in the gaseous regime $\gamma_2/n_2=100$ located at $\alpha=1/3$, as a function of its eccentricity $e_2$ and the binary mass ratio $m_1/m_0$. The white curve represents the perfect synchronous state $\langle \Omega_2 \rangle = n_2$, which divides the map in a region where the stationary spins are expected to be super-synchronous  (above), and another one where sub-synchronous spins are predicted (behind). With this reference, we can observe that the size of the sub-synchronous region is proportional to the normalized secondary mass, and tends to disappear for low secondary companion. In competition with $m_1/m_0$, as we noticed in \cite{Zoppetti2019,Zoppetti2020} and similar to that expected with the 2BP, the planetary eccentricity $e_2$ leads to more rapid stationary spins. 

We have included in the Figure \ref{fig:keplers} the nominal location in the diagram of the {\it{Kepler}} CB planets observed very close to the binary and with an estimated mass which allows us to infer that they have an important gaseous component ($m_2 \gtrsim m_{Neptune}$): Kepler-16 \citep{Doyle2011}, Kepler-35 \citep{Welsh2012}, Kepler-38 \citep{Orosz2012}, Kepler-47b \citep{Orosz2012b}, Kepler-64 \citep{Schwamb2013}, Kepler-413 \citep{Kostov2014} and Kepler-1661 \citep{Socia2020}. If we assume that all these CB planets have an associated relaxation factor that satisfy $\gamma_2 >> n_2$ and that, in turn, these viscosities allowed them to have already reached their equilibrium state, so the expected stationary spins for them is typically sub-synchronous. The exceptional case is Kepler-413, an eccentric CB planet orbiting a binary with two comparable-mass central stars, while Kepler-64 is expected to be very close to the perfect synchronous state. However, as we mention in Section \ref{sec:nume}, the amount of sub-synchronous shift is much less important than the expected with the CTL tide model built in our previous works \cite{Zoppetti2019,Zoppetti2020}. The reasons of this difference will be explained in detail in the following Section.



\section{Revisiting the rotation of a CB planet in the CTL model}
\label{sec:Mignard}


In the limit of gaseous bodies (i.e. $\gamma_2 >> n_1,n_2$), the stationary spin equation (\ref{eq:ome_st}) expanded up to 2nd-order in $\alpha$ becomes

\begin{eqnarray}
\label{eq:ome_st_mig}
 \langle \Omega_2 \rangle_{st}^{gas} &=& n_2 + 6 n_2 e_2^2 + 5 n_2 e_2^2 \mathcal{M}_2 \alpha^2. 
\end{eqnarray}
We notice that this expression is different than the one we obtained in equation (22) of \cite{Zoppetti2019}, using the classic CTL tidal model, based on the expressions of the direct torques of \cite{Mignard1979}. 
This Section is dedicated to investigate the origin of this discrepancy and comparing these two different tidal models (i.e. CTL and creep) in the framework of the 3BP and, thus, to obtain conclusions about the advantages and disadvantages of each one, in different contexts.

In Section (2.1) of \cite{Zoppetti2019} we used a simplified model to test the role of the cross tides in the 3BP with the CTL tidal model of \cite{Mignard1979}. We considered only one extended mass $m_0$ under tidal interaction of only one companion $m_1$ and used a third mass $m_2$ only as a test particle, to evaluate the secular contribution of the cross torque, neglecting the gravitational perturbations of $m_1$. We used the usual astrocentric coordinate system, centered on the extended mass.
In that simplified system we found, employing numerical integrations and also analytical calculations, that the secular net contribution of the cross torques on the test particle were null. Then, we extrapolated that result to the general 3-extended body problem in a Jacobian frame and neglected the cross forces and tides of Mignard expressions. In what follows, we reformulate and improve the CTL model developed in \cite{Zoppetti2019} with the inclusion of the cross tides, and compare with our previous results.



\begin{figure}
\centering
\includegraphics[width=0.8\columnwidth]{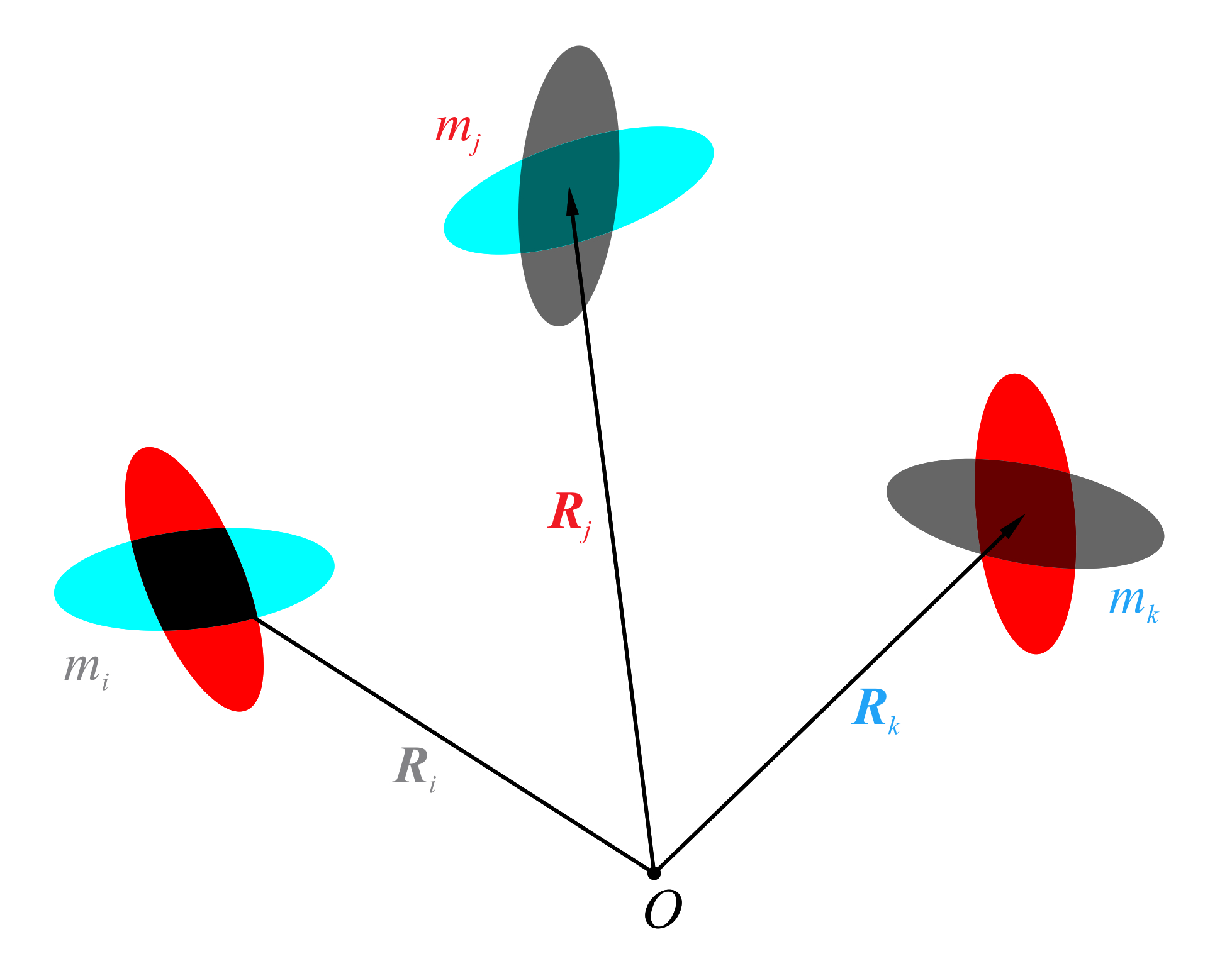}
\caption{Sketch of the 3BP considering all the tidal interactions between pairs of bodies. The grey ellipsoids are induced on the companions due to the presence of $m_i$, the red ellipsoids due to the presence of $m_j$ and the blue ellipsoids are due to the presence of $m_k$.}
\label{6elip}
\end{figure}

Once again, we consider the 3BP composed of $m_i$, $m_j$ and $m_k$ represented in Figure \ref{6elip}, where between each pair of bodies we compute the complete tidal interaction, in addition to the gravitational forces. As a result of this consideration of the tides, due to the presence of each body, let's say $m_i$, one ellipsoid is generated on $m_j$ and one ellipsoid on $m_k$, both represented as grey ellipsoids on each respective mass of Figure \ref{6elip}.  

Let us denote as ${\bf{F}}_{i,j}^k$ to the tidal force applied on $m_i$ due to the ellipsoid that the presence of $m_k$ generated on $m_j$. In Figure \ref{6elip}, ${\bf{F}}_{i,j}^k$ corresponds to the force applied on $m_i$ due to the blue ellipsoid generated on $m_j$. Thus, if we take into account all the action forces applied on $m_i$, we must consider the tides caused by the two ellipsoids that $m_i$ itself generates on its companions (grey ellipsoids) plus the tides due to the cross ellipsoids (blue ellipsoid on $m_j$ and red ellipsoid on $m_k$). On the other hand, to complete the force diagram, we also need to consider the reaction forces that $m_i$ exerts on its two companions, due to the ellipsoids generated on itself (blue and red ellipsoids on $m_i$). As a result, the total tidal force applied on $m_i$ is given by

\begin{eqnarray}
\label{eq:f8}
{\bf{F}}_{i} = \underbrace{( {\bf{F}}_{i,j}^i + {\bf{F}}_{i,j}^k + {\bf{F}}_{i,k}^i + {\bf{F}}_{i,k}^j )}_{action \ tides}
- \underbrace{( {\bf{F}}_{j,i}^j + {\bf{F}}_{j,i}^k + {\bf{F}}_{k,i}^k + {\bf{F}}_{k,i}^j )}_{reaction \ tides}
\end{eqnarray}

We note that the forces with three different indexes in equation (\ref{eq:f8}) correspond to the cross tides, which were neglected in \cite{Zoppetti2019}, while the forces with two repeated indexes correspond to the direct tides, which we did consider in our previous work.

If we adopt the classical formalism of \cite{Mignard1979} to model the tidal interactions between each pair of bodies with, the explicit expression of the force ${\bf{F}}_{i,j}^i$ is expanded up to first order in the time lag $\Delta t_j$ as
\begin{equation}
{\bf{F}}_{i,j}^i = {\bf{F}}_{i,j}^{k,(0)} + \Delta t_j {\bf{F}}_{i,j}^{k,(1)},
\end{equation}
where (see \cite{Zoppetti2019})
\begin{eqnarray}
{\bf{F}}_{i,j}^{k,(0)} = \frac{3}{2} \mathcal{G} \mathcal{R}_j^5 k_{2,j} \frac{m_i}{\Delta_{ij}^5} \frac{m_k}{\Delta_{kj}^5} \bigg[ 2({\bf\Delta}_{ij} \cdot {\bf\Delta}_{kj}) {\bf\Delta}_{kj} + \nonumber \\ (\Delta_{kj}^2 - \frac{5}{\Delta_{ij}^2} {({\bf\Delta}_{ij} \cdot {\bf\Delta}_{kj})}^2) {\bf\Delta}_{ij}\bigg], 
\end{eqnarray}
\begin{eqnarray}
\label{eq:fgral}
{\bf{F}}_{i,j}^{k,(1)} = 3 \mathcal{G} \mathcal{R}_j^5 k_{2,j}  \frac{m_i}{\Delta_{ij}^5} \frac{m_k}{\Delta_{kj}^5} \bigg[ \frac{({\bf\Delta}_{kj} \cdot \dot{{\bf\Delta}_{kj}})}{\Delta_{kj}^2} \bigg( 5 {\bf\Delta}_{kj} ({\bf\Delta}_{ij}\cdot{\bf\Delta}_{kj})-\Delta_{kj}^2 {\bf\Delta}_{ij} \bigg)    \nonumber \\
- [{\bf\Delta}_{kj} \cdot ({\bf\Omega}_j \times {\bf\Delta}_{ij}) + {\bf\Delta}_{ij} \cdot \dot{{\bf\Delta}_{kj}}]{\bf\Delta}_{kj} - [{\bf\Delta}_{kj} \times {\bf\Omega}_j + \dot{{\bf\Delta}_{kj}}] ({\bf\Delta}_{ij} \cdot {\bf\Delta}_{kj})  \nonumber \\
+ \frac{5}{\Delta_{ij}^2} {\bf\Delta}_{ij} \bigg( ({\bf\Delta}_{ij} \cdot {\bf\Delta}_{kj}) [ {\bf\Delta}_{kj} \cdot ({\bf\Omega}_j \times {\bf\Delta}_{ij} ) + {\bf\Delta}_{ij} \cdot \dot{{\bf\Delta}_{kj}} ]  \nonumber \\
- \frac{({\bf\Delta}_{kj} \cdot \dot{{\bf\Delta}_{kj}})}{2 \Delta_{kj}^2}(5 {({\bf\Delta}_{ij} \cdot {\bf\Delta}_{kj})}^2- \Delta_{ij}^2\Delta_{kj}^2) \bigg) \bigg],
\end{eqnarray}
with $k_{2,j}$ the second degree Love number of $m_j$. 

The conservation of the total angular momentum of the system imposes that the total orbital angular momentum ${\bf{L}}_{orb}$ must satisfy
\begin{equation}
\label{eq:conser}
\sum_{i=0}^2 C_i \dot{\bf{\Omega}_i} = -\dot{\bf{L}}_{orb},    
\end{equation}
where the variation of the total orbital momentum can be written in terms of the relative positions as \citep{Zoppetti2019}
\begin{eqnarray}
\dot{\bf{L}}_{orb} = 
{\bf\Delta}_{j,i} \times ({\bf{F}}_{j,i}^j+{\bf{F}}_{j,i}^k) + 
{\bf\Delta}_{k,i} \times ({\bf{F}}_{k,i}^k+{\bf{F}}_{k,i}^j) + \\
{\bf\Delta}_{i,j} \times ({\bf{F}}_{i,j}^i+{\bf{F}}_{i,j}^k) +
{\bf\Delta}_{k,j} \times ({\bf{F}}_{k,j}^k+{\bf{F}}_{k,j}^i) + \nonumber \\
{\bf\Delta}_{i,k} \times ({\bf{F}}_{i,k}^i+{\bf{F}}_{i,k}^j) +
{\bf\Delta}_{j,k} \times ({\bf{F}}_{j,k}^j+{\bf{F}}_{j,k}^i) \nonumber
\end{eqnarray}

As discussed in Section \ref{sec:ana}, equation (\ref{eq:conser}) may be decoupled if we assume that the variation in the spin angular momentum of the body $m_j$ is only due to the terms in the equation associated with its own deformation (see \cite{Zoppetti2019}). Using this boundary condition, we have

\begin{eqnarray}
\label{eq:t_migna}
C_i \dot{\bf{\Omega}}_i = - \bigg[ \underbrace{ ( {\bf\Delta}_{ji} \times  {\bf{F}}_{ji}^j + {\bf\Delta}_{ki} \times  {\bf{F}}_{ki}^k ) }_{direct \ torques}  
+ \underbrace{ ( {\bf\Delta}_{ji} \times  {\bf{F}}_{ji}^k + {\bf\Delta}_{ki} \times  {\bf{F}}_{ki}^j )}_{cross \ torques} \bigg] 
\end{eqnarray}
Again, we mention that the rotational evolution descripted by equation (15) of \cite{Zoppetti2019}, only considers the direct torques, i.e. the first two terms in equation (\ref{eq:t_migna}). 
\begin{figure}[h!]
\centering
\includegraphics*[width=0.9\columnwidth]{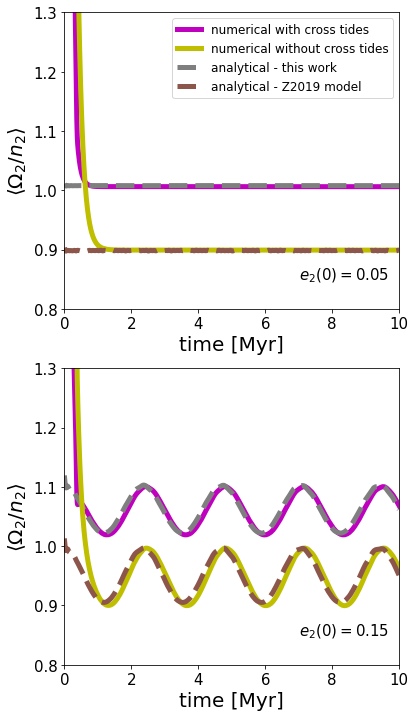}
\caption{ Rotational evolution of a CB planet with the CTL model of \cite{Mignard1979} (equation (\ref{eq:t_migna})). The initial conditions were taken from Table \ref{tab1}, with the exception of the initial planetary eccentricity ($e_2(0)=0.05$ in the upper panel and $e_2(0)=0.15$ in the bottom panel).  Full curves indicate the results of filtered numerical integration: excluding the cross tides (in yellow) and considering the full expression  (in magenta). Dashed curves represents the results of different analytical models: our previous CTL tidal model \citep{Zoppetti2019} (in brown) and the results of the analytical creep tidal model presented in equation (\ref{eq:ome_st}) of this work, expanded up to 2-order in $\alpha$ and considering the gaseous limit ($\gamma_2 >> n_1,n_2$), are shown in grey. } 
\label{figmigna}
\end{figure}

In Figure \ref{figmigna}, we compare the different rotational evolution that results from the CTL tidal model of \cite{Mignard1979} in the 3BP, by considering the full expression in equation (\ref{eq:t_migna}) and the version constructed neglecting the cross tides. The initial conditions were taken from Table \ref{tab1}, except for the initial planetary eccentricity which was varied: $e_2(0)=0.05$ in the upper panel and $e_2(0)=0.15$ in the bottom panel. In full curves we show the results of digitally filtered numerical integrations of equation (\ref{eq:t_migna}), considering the full expression of the torque equation (in magenta) and neglecting the cross tides (in yellow). We have also included in this simulations the punctual gravitational interactions between the bodies, to compare with our previous simulations of \cite{Zoppetti2019}.

The results obtained with different analytical models are shown with dashed curves in Figure \ref{figmigna}: with brown curves we represent the results of the analytical CTL tidal model of \cite{Zoppetti2019} (equation (22)) and with grey curves we represent the results of the analytical creep tidal model presented in this work (equation (\ref{eq:ome_st})) expanded up to 2-order in $\alpha$, in the gaseous case ($\gamma_2 >> n_1,n_2$).

The comparison between the numerical integrations of the torque equation (\ref{eq:t_migna}) without cross tides and the analytical results of the CTL model of \cite{Zoppetti2019} are analogous to the one presented in the right panels of Figure 7 of that work. However, here we adopt a tidal parameter $Q_2' = 3/(2 n_2 k_{2,2} \Delta t_2)$ that has been chosen in order to the associated relaxation factor satisfy $\gamma_2 = 100n_2$ (see \cite{FerrazMello2013}) and, thus, the rotational evolution timescales be similar to the one presented in the magenta simulation of Figure \ref{fig:caidas}.

We first notice that the stationary solution of $\langle \Omega_2 \rangle$ exhibits long-term periodic oscillations, specially for the initially eccentric case. As mention in \cite{Zoppetti2019}, this is due to the secular variation induced on the planetary eccentricity by the gravitational punctual interactions.

Regarding the comparison with the numerical simulations, we can notice in both panels of Figure \ref{figmigna}, that the analytical model presented in \cite{Zoppetti2019} (brown curves) fits very well the stationary value of the planetary rotation, when we neglect the cross tides in the torque equation. This is an expected result since, in the construction of that model, we took into account only the direct tidal forces and not the general expression of equation (\ref{eq:fgral}).

The comparison of the numerical integration for the full torque equation with the analytical results obtained with the creep model (in the gaseous limit), reveals two very important facts. On one hand, from the proximity of the magenta curves with the grey curves, we show that the results obtained for the rotational stationary state with the CTL tidal model of \cite{Mignard1979} are the same that the one obtained with the creep model in the limit of very large relaxation factor ($\gamma_2 >> n_1,n_2$). This result was checked in a completely analytical calculation, by expanding the position and velocity vectors of equation (\ref{eq:t_migna}) in $\alpha$ and the eccentricities, and then averaging with respect to both mean longitudes, with the help of an algebraic manipulator. 

On the other hand, the difference between the magenta and yellow curves of Figure \ref{figmigna}, shows that, unlike what we assumed in \cite{Zoppetti2019} from analyzing the simple case in which  we considered only one extended mass and a reference frame in solidarity with it, the cross tides are not negligible in the general self-consistent problem of three extended masses. However, the physical effect of the tides due to the presence that a binary (instead a single star) exerts on an external planet is the same as we reported in our previous works: it shifts the rotational stationary solution to lower spin values, for an amount that is proportional to the semimajor axes ratio $\alpha$ and the secondary mass $m_1$. The no-cross tides model developed in \cite{Zoppetti2019,Zoppetti2020} simply overestimate this effect.

\section{Summary and discussion}

In this work, we present a model for treating the tides in the general 3BP, in the case in which all the bodies are considered extended and tidally interacting. In particular, we consider the planar case where all the bodies have spins vectors pointing in a direction perpendicular to the orbital plane. We use the creep tide formalism first presented in \cite{FerrazMello2013} to compute the tidal forces and torques, and apply it to study the rotational evolution of a circumbinary planet. The polar flattening due to the rotation of the bodies is also naturally taken into account in the theory \citep{Folonier2018}. 

First, we present the creep differential equations that govern the rotational evolution of the extended bodies, in the framework of the 3BP. 
We show that, on each body, the tidal interactions due to the presence of the two companions can be modelled by an equivalent unique interaction. This allows the formalism to be extended to the general {\it N}-bodies tidally interacting case, where the difficulty lies in finding the parameters of the equivalent interaction. As in the 2BP, the physical quantities that characterize the rotational evolution of each body in this model are the spin velocities, the equivalent lag-angles, tidal flattenings and polar flattenings. 

We then apply the model to study the rotational evolution of a circumbinary planet, using as a working example a Kepler-38 like system \citep{Orosz2012}. To do this, we take advantage of the adiabatic nature of the rotational respect to the orbital evolution in this problem, which let us assume as fixed the orbital elements with the exception of the fast angles.

Using direct numerical integrations, we find that, for planets in low eccentric orbits and viscosities in the range of the ones estimated for Solar System planets by \cite{FerrazMello2013}, the more probably final stationary state is the 1:1 spin-orbit resonance, with the equivalent lag-angle oscillating around 0. The required timescales for reaching the stationary state depends on the relaxation factor of the bodies:
taking as a reference the relaxation factors estimated for \cite{FerrazMello2013}, we expect that CB stiff planets in a Kepler-38 like system synchronize its spins in timescales much lower than the age of the system, while this is not necessary the case for planets with a large gaseous component. However, the great uncertainty about the internal structure of exoplanets and, even more, about their associated viscosities inhibit an accurate estimate of its rotational evolution timescale.

The high probability of capture in the 1:1 spin-orbit resonance, for low eccentric CB planets with arbitrary viscosities, motivated us to study this particular solution in more detail. With this aim, we search for a high-order analytical expression for the mean solution of the rotational stationary state. 
The extreme proximity to their central binary to which Kepler's CB planets have been observed, in addition to the strong dependence of tidal effects with the relative distance between the bodies, forced us to develop a high-order model in semimajor axes ratio. As a result of this, our solution has shown to be very precise in predicting the full numerical integrations, for arbitrary planetary viscosity and very high $\alpha$-values, even inside the expected gravitationally instable region.  

From a practical point of view, when studying long-term tidal evolution, having a secular approximation of the stationary rotational solution will considerably accelerate the numerical simulation of the full system, avoiding having to integrate the rotational evolution and, instead, introducing an analytical recipe for the equilibrium state. This semi-analytic method is particularly useful in the case of working with gaseous bodies with large relaxation factors, for which the creep equation becomes particularly complex to solve efficiently.

As an application of our analytical model, we consider the {\it{Kepler}} CB planets observed close to the central binary and with an expected large gaseous component. If we assume that all these CB planets have associated viscosities that let them to have reached their rotational steady state, our model predicts that their typical rotation speeds are sub-synchronous. This result is mainly due to the presence of a central binary (instead of a single star) and is predicted by our model for very low eccentric CB planets close to a binary with a high mass ratio. However, the stationary rotation for gaseous CB planets predicted by the creep tide theory presented here is different than the one expected with the CTL model presented in \cite{Zoppetti2019,Zoppetti2020}. 

To explain this discrepancy, we finally discuss the results obtained for the stationary spin of a CB planet, with the creep tide theory (in the gaseous limit) and compare with those obtained using the CTL tidal model in \cite{Zoppetti2019,Zoppetti2020}. We show that both model predict exactly the same solution as long as the cross torques are taken into account in the CTL formalism. These torques had been neglected in our previous works, from the study of the classical case of a single extended body with a system of coordinates attached to it. However, for the general case of extended 3BP and an arbitrary reference frame, this is not necessarily true. Actually, one of the main advantage of the creep tide theory over CTL model is the treatment of cross torques. While in the CTL model the general expressions for the tidal force and torques are very complex and only valid for small and constant time-lags, in the creep theory these quantities are calculated from the equilibrium figures where the cross torques are naturally taken into account. 

In addition to this, the CTL model seems to be contained within the creep tide theory, for the case of relaxation factors much higher than the fundamental frequencies of the system. Moreover, in the creep tide formalism the lag-angle is not an external {\it{ad-hoc}} quantity which needs to be small and constant, but can take large values and its temporal evolution is solved from a differential equation. Finally, among other of its most interesting advantages, we name that the calculation of an equivalent equilibrium figure allows not only to take into account the cross torques in a natural way but also to incorporate some additional physical phenomena such as polar flattening.  As a result, the theory also provide the evolution of the geometric shape of the interacting bodies.

Among its main limitations, we found some difficulties in numerically solving the creep equation for gaseous bodies with high relaxation factors in an efficient way. This means that, when studying long-term tidal evolution, some analytical recipes need to be incorporated to accelerate the numerical integrations, such as the mean values that describe the rotational evolution presented in this work.
On the other hand, as noticed in \cite{FerrazMello2015b}, the creep theory is analog to those which consider as Maxwell bodies to the extended masses \citep{Correia2014}, but without the elastic component of the tide.  More complex rheologies such as Sundberg-Cooper model \citep{Sundberg2010} may be neccesary to reproduce many of the intricacies seen in the deformation of real materials \citep[e.g.][]{Renaud2018}.

\section*{Appendix: 6th-order terms for the mean stationary solution}
\label{ap:1}

The non-eccentric terms expanded up to 6th-order in $\alpha$ of expressions (\ref{eq:ome_st}), (\ref{eq:delta_st}), (\ref{eq:er_st}) and (\ref{eq:ez_st}) are respectively given by

\begin{eqnarray}
 \langle \Omega_2\rangle_{o6}=&-\frac{5 \mathcal{M}_2 (n_1-n_2)\gamma_2^2\alpha^6 }{32 \left(\gamma_2^2+(n_1-n_2)^2\right)
 \left(\gamma_2^2+(2n_1-2n_2)^2\right) \left(\gamma_2^2+(3n_1-3n_2)^2\right)} \\
\times& \bigg[5 \ \bigg(121\gamma_2^4+805\gamma_2^2(n_1-n_2)^2+744(n_1-n_2)^4\bigg) \mathcal{M}_4 \nonumber \\
\qquad& - \bigg(729\gamma_2^4+5265\gamma_2^2(n_1-n_2)^2 +4836(n_1-n_2)^4 \bigg) \mathcal{M}_2^2\nonumber \bigg]
\end{eqnarray}
\begin{eqnarray}
 \langle \delta_2\rangle_{o6} = \frac{\langle \Omega_2\rangle_{o6}}{\gamma_2}
\end{eqnarray}
\begin{eqnarray}
\frac{\langle\mathcal{E}^\rho_2\rangle_{o6}}{\overline{\epsilon}^\rho_2}=&\frac{5M_2 \alpha^6}{256\left(\gamma_2^2+(n_1-n_2)^2\right) \left(\gamma_2^2+4 (n_1-n_2)^2\right) \left(\gamma_2^2+9 (n_1-n_2)^2\right)} \\
\times& \Big[(105\mathcal{F}_1+776\mathcal{F}_2)\gamma_2^6  +70 (21\mathcal{F}_1+76\mathcal{F}_2)\gamma_2^4(n_1-n_2)^2 \nonumber \\
\qquad&+(5145\mathcal{F}_1+5024\mathcal{F}_2)\gamma_2^2(n_1-n_2)^4+378\mathcal{F}_1 (n_1-n_2)^6 \Big]\nonumber
\end{eqnarray}
\begin{eqnarray}
\frac{\langle {\mathcal E}^z_2 \rangle_{o6}}{\overline{\epsilon}^M_2} = \frac{2\langle\Omega_2\rangle_{o6}}{n_2} +\frac{\overline{\epsilon}^\rho_2}{\overline{\epsilon}^M_2} \frac{1225}{512} \mathcal{M}_2 \mathcal{F}_1 \alpha^6
\end{eqnarray}
where
\begin{eqnarray}
\mathcal{F}_1 = \frac{1+\beta^5}{(1+\beta)^5} \ \ &;& \ \ 
\mathcal{F}_2 = \frac{\beta (1-\beta)^2}{(1+\beta)^4}.
\end{eqnarray}

\begin{acknowledgements}
     This research was funded by CONICET, SECYT/UNC, FONCYT and FAPESP (Grant 2016/20189-9 and 2016/13750-6. The authors also wish to thank the anonymous referee for helpful suggestions that greatly improved this work)
\end{acknowledgements}

%
%

\end{document}